# Using Assembly Language for Creating Games

Haris Turkmanović, David Vukoje, Aleksandra Lekić, *Member*, *IEEE,* and Milan Prokin

*Abstract*— The aim of this paper is to demonstrate some interesting and useful approaches for writing a program in the assembly language. In order to demonstrate the possibilities of the assembly language, a project called "Arkanoid" was created. This project is written in assembly language and it presents few interesting algorithms. Assembly language, which is used for designing the game is x86 Assembly language, which produces object code for the x86 class of processors. As a working environment is chosen Visual Studio 2015, because it gives the useful tools for debugging and testing of the created software (game). Execution of the program results in a "Arkanoid" game, placed in Windows OS Console.

## I. INTRODUCTION

THE assembly language is a low-level programming language, which offers an efficient way for writing programs. Under the efficient, we assume that the programmer has a complete control over the code organization. This is possible because there is no compiler which translates and organizes the code. Thus, writing the program and the code organization is a full responsibility of the programmer. The programmer has an access to each memory address and the full control of each byte in memory [1-3]. This feature made assembly a very popular in writing fast interrupt procedures in the world of embedded systems over the years [4-6]. Writing some parts of the "fast procedures" or functions in the assembly is especially used for digital signal processors programming [7].

The fact that developer has the full control of each byte stored in memory leads to the conclusion that the problems in the program can be produced if developer does not have a clear development plan and code organization. There are two reasons for introducing high-level programming languages. The first reason was to relieve the programmer of responsibility in terms of organizing memory and the second key factor was to put code abstraction on a higher level (ex. structure in C is similar to a class in C++ where class offers a higher level of code abstraction) [8].

In our project, the code organization is done using x86 assembly language low-level abstraction method (like structures, procedure, macros, etc.), which makes algorithms safer and easier readable [1, 3]. All algorithms written in the assembly language are grouped in one project named "Arkanoid". The "Arkanoid" is a game which demonstrates:

Haris Turkmanović, David Vukoje, Aleksandra Lekić and Milan Prokin are with the School of Electrical Engineering, University of Belgrade, 73 Bulevar kralja Aleksandra, 11020 Belgrade, Serbia (e-mail: th140516d@student.etf.bg.ac.rs, vd140541d@student.etf.bg.ac.rs, lekic.aleksandra@etf.bg.ac.rs, proka@etf.bg.ac.rs).

- The correct operation of the implemented algorithms;
- Creating interfaces in Windows operating system console;
- The possibilities of the assembly language (macros, procedure, structure, etc.).

The paper is organized in five sections. In the section II the organization of the assembly programming language is shortly presented. Section III provides description of the game "Arkanoid" and of the most interesting concepts and procedures used. In the section IV is described the graphical interface for the game, while the section V consists of the specific algorithms' descriptions. Concluding remarks are given in section VI.

## II. USING ASSEMBLY LANGUAGE

Before we start explaining the concrete solutions of the "Arkanoid" implementation, it is important to give an explanation of the assembly basic concepts. Here we introduce some concrete assembler assumptions implemented in the program "Arkanoid".

The assembly language translates directly a line of code to the machine code [1]. Thus, it is connected with processor architecture and the code is very dependent of the architecture used. For that matter it is necessary to get acquainted with the actual processor architecture where the code will be executed. It implies that the interpretation of all registers, the ways of addressing, the length of the processor word, instruction set, etc. has to be known.

Project "Arkanoid" is written on Intel 32bit I5 processor core which used by standard home PC. The aim was to demonstrate the use of instruction set that Intel created for 8086 architecture [1, 3]. Intel processor made it possible because programs written for an older Intel architecture are completely compatible with the new Intel architecture. That is very useful, because now days PC processors have more peripherals and the work with 8086 assembly language gives more pleasure and some more interesting features can be implemented. In the Fig. 1 is depicted Intel 8086

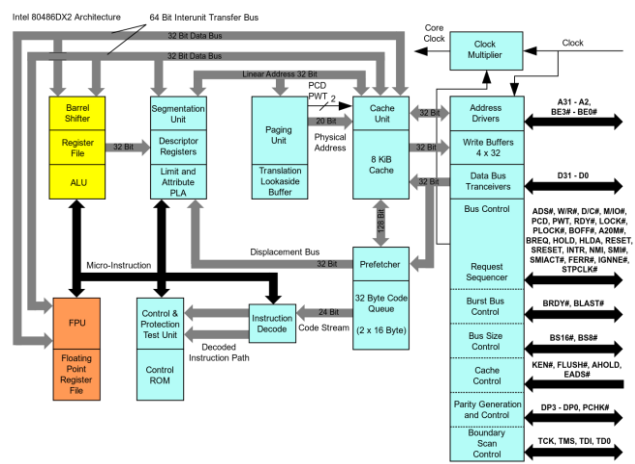

Fig. 1. Intel 80486 architecture.

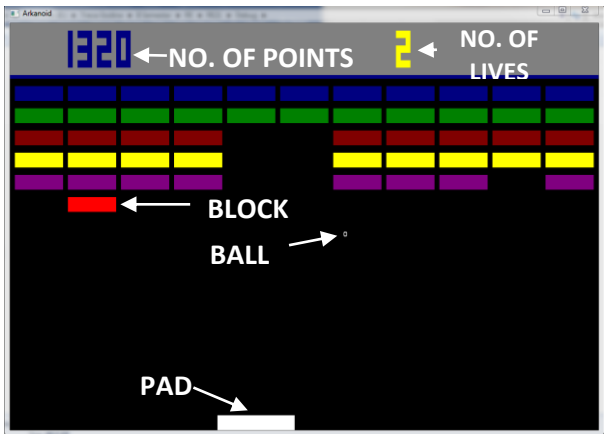

Fig. 2. Game interface.

architecture used for which the project is realized.

## III. IMPLEMENTATION OF THE "ARKANOID"

### A. "Arkanoid" features

The game "Arkanoid" satisfies following requirements:
- Graphic interface which contains the ball, pad, and blocks;
- The ball can hit border game area (console border, pad, currently presented blocks, etc.). If the ball hits the bottom of the console, the game is over;
- When the ball hits left, top or right console border, blocks or pad, it must be rebound at the certain angle;
- Each block has a color and each color carries a certain number of points;
- When the ball hits the block, the block disappears and the current player gains a number of points. This number of points represents the value of the disappearing block's color;
- Pad, which does not allow the ball to hit the bottom of the console, can be moved by a player left or right;
- During the game, the current number of points and lives is printed. Anytime during the game, user can cancel it by pressing ESC keyboard button;
- When the first row blocks are hit by the ball for the first time, blocks only change color. After the second hit the first row block with the changed color falls. If the pad catches falling block, a number of points is gained which presents a doubled block value.

Game graphic interface is show in the Fig. 2.

### B. "Arkanoid" structure

All procedures created to handle the game requirements can be divided into three groups of procedures: graphic procedures, user command procedures and algorithms that control program flow. This is illustrated in the Fig. 3.

Graphic procedures handle graphics object like pad, blocks, ball, points, strings (number of points, number of lives, etc.). User command procedures handle command received by the user from the keyboard like left navigation button, right navigation button, ESC button, etc. Control algorithms are in charge of data processing. The flow of the program is further controlled by processing result.

### C. I/O Procedures

Input or Output procedures are special type procedures that can't work independently, which implies that their work requires interaction with the operating system function. The operating system function call will be explained in detail in the following text.

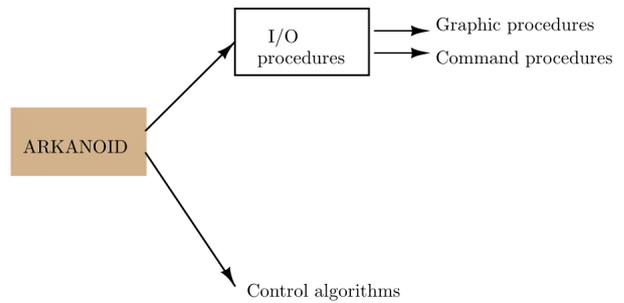

Fig. 3. Illustration of Arkanoid Structure.

Let's assume that a programmer wants to read pressed button on the PC keyboard. Frist, the programmer needs to know the architecture of the peripheral named "Keyboard". After that, he needs to know how processor communicates with the keyboard and how the memory maps peripherals. Even if the programmer knew the previous steps and if his code works on that particular processor with the keyboard, his code wouldn't work on other machines. There are lot of reasons why that code wouldn't work on other PCs (different memory organization, different peripherals memory map, different processor, etc.) [1, 3].

Thus, operating system functions (OSF) abstract access to the hardware. This means that if the programmer wants to access hardware, he just needs to call the operating system function with the proper arguments and as a result, he would receive response from the hardware (the key is pressed, the string is written on the console, etc.). OSF is something like interface to hardware.

Windows OS gives to the programmer library OSF to communicate with I/O peripheries [3]. All OSF are stored in *kernel32.lib*. This is a very useful library for making an assembly application. Some of the functions from this library used in this project are *WriteConsoleOutput*, *WriteConsoleOutputChar*, *SetConsoleWindowInfo*, *SetConsoleCursorPosition*, etc. In the Fig. 4 is depicted an access of the application to the hardware through OSF function.

## IV. GRAPHIC INTERFACE

Before describing how the graphic interface of the game is implemented, it is necessary to explain how to work with the Windows console.

The purpose of the console is to communicate with the user. The console can be used to display data or to take data from the user. One of console parts is a *console screen buffer*. That is a two-dimensional array of characters which has ASCII value, background and foreground color. In other words, the console can be imagined as an array of rectangles as depicted in Fig. 5. Each rectangle can adjust its background color and in the each rectangle can be written a specific character. Specific character is in the ASCII code ant it is displayed in that particular rectangle. These two properties of console rectangle are grouped and form structure named "CHAR_INFO" which has two elements: *ASCII code* and *attributes* [1].

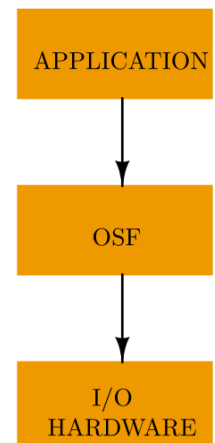

Fig. 4. Access to the hardware from the perspective of the application.

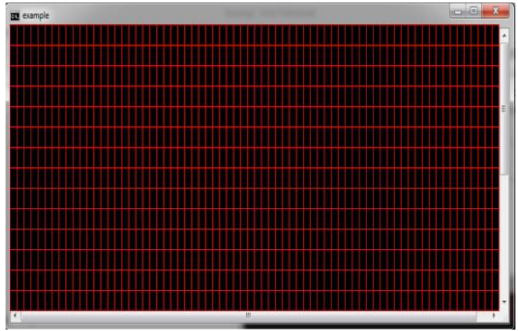

Fig. 5. Windows console with show rectangle network.

Access to the console screen buffer is abstracted with *kernel32.lib* procedures. Commonly used procedure for writing to the console is *WriteConsoleOutput* which allows console printing of the CHAR_INFO structure objects. *WriteConsoleOutput* function allows printing of the two-dimensional array consisting of CHAR_INFO structure objects to the particular position over the console, which is a key feature in creating graphic objects of the game.

The graphical interface of the game "Arkanoid" consists of two parts. One part is "Welcome window interface" and the second part is "Game interface". Both interfaces contain an important graphic concept named "Big strings".

### A. Welcome window interface

Strings present groups of characters. In the "Arkanoid" we have "Big strings" like a group of "Big chars" (letters/symbols). Chars are ASCII chars whose main feature is to present ASCII values. The main property of the Big char is "Char bitmap" which represents a group of ASCII characters (usually have value 219 which is ASCII code for rectangle), which gives a graphic representation of the desired symbol's letter. Each ASCII value has corresponding "Big char bitmap". This concept is illustrated in the Fig. 6.

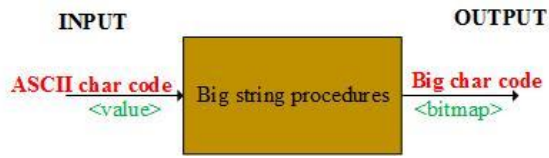

Fig. 6. Concept of BigString idea.

Basic building elements for "Char bitmap" are two CHAR_INFO structure elements, which represent full and empty rectangle in the console. Full rectangle forms shape of "Big chars" (symbols). Empty rectangle forms "Big chars" background. Combining the empty and the full rectangle we can create random shapes. This is very useful for creating different sized arbitrary forms and allows us bigger creativity.

"Char bitmap" is implemented as a 2D array of CHAR_INFO elements, which abstracts empty and full rectangle. The project supports three different sizes of "Char bitmap": 3x5, 5x7, 9x11, which are implemented in 3 different include files *fontemplate3x5.inc*, *fontemplate5x7.inc*, *fontemplate9x11.inc*. Part of the contents of one of these files is show in the Fig. 7. Each symbol bitmap has a unique ID which makes the access to that bitmap much easier. If bitmap represents a symbol of ASCII table char, bitmap ID is equal to the character ASCII value. Otherwise, bitmap has an ID that is greater than 255. This concept makes usage of "Big char" more flexible.

If the user wants to add his own symbol, he needs to create symbol bitmap in the include file *fontemplate.inc* and to join ID value to that symbol. One of the custom created symbols is illustrated in Fig. 8. That symbol represents School of the Electrical engineering University of Belgrade.

Working with "Big string" is very easy because the access to bitmap is abstracted through the procedures which make work with "Big string" much intuitive. There are two main procedures: *WriteBigString* and *WriteLetter*. *WriteBigString* is used when the user wants to print ASCII string to the console in a form of "Big String" and *WriteLetters* is a procedure which allows to print a particular bitmap to the console screen. All of these procedures allow printing on a particular position in the console window (Fig. 5). This property makes the idea of "Big String" more useful and makes working with "Big string" more flexible. As a result of using "Big string" is gotten "Welcome window interface" shown in the Fig. 9.

### B. Game interface

The game interface depicted in the Fig. 2 can be separated into two parts. One part is located at the top of the console window (gray background). That part of the game interface is named "Info part". It is used to show current game info like a number of points or a number of remaining lives. "Info part" is created using "Big string" procedures. Below "Info part" is located "Game part", where are graphics parts of the game like *blocks*, *pad*, and *ball*.

*Block* is a part of the game interface which has properties like color, number of points, position on the console window, etc. Blocks are objects of the structure named "CELL" which stored the information important for blocks like value, color, area coordinates, etc. One of the important fields of CELL structure is "AREA_Position", which

```
Font_Block        = 219
Font_Color        = Red
Font_Backround    = 80h

Point             TEXTEQU    <<Font_Block,Font_Color>>
None              TEXTEQU    <<0,Font_Backround>>

.data
Letter_Size1  COORD <3,5>

;This is template for SPACE char
Delete_font10   _CHAR_INFO None,None,None
Delete_font11   _CHAR_INFO None,None,None
Delete_font12   _CHAR_INFO None,None,None
Delete_font13   _CHAR_INFO None,None,None
Delete_font14   _CHAR_INFO None,None,None

;This is template for Digit 1
Number_One      _CHAR_INFO None,None,Point
Number_One1     _CHAR_INFO None,None,Point
Number_One2     _CHAR_INFO None,none,Point
Number_One3     _CHAR_INFO None,None,Point
Number_One4     _CHAR_INFO None,None,Point
```

Fig. 7. Part of file named fonttemplate3x5.inc which illustrate bitmap for digit 0 and 1.

```
Font_Block       = 219
Font_Color       = Red
Font_Backround   = 80h
Point    TEXTEQU <<Font_Block,Font_Color>>
None     TEXTEQU <<0,Font_Backround>>
.data
Letter_Size3 COORD <9,13>
ETF_logo    _CHAR_INFO Point,Point,Point,Point,Point,Point,Point,Point,Point
ETF_logo1   _CHAR_INFO Point,None,None,None,None,None,None,None,Point
ETF_logo2   _CHAR_INFO Point,None,Point,Point,Point,Point,Point,None,Point
ETF_logo3   _CHAR_INFO Point,None,Point,None,None,None,Point,None,Point
ETF_logo4   _CHAR_INFO Point,None,Point,None,Point,None,Point,None,Point
ETF_logo5   _CHAR_INFO Point,None,Point,None,Point,None,Point,None,Point
ETF_logo6   _CHAR_INFO Point,Point,Point,Point,Point,Point,Point,Point,Point
ETF_logo7   _CHAR_INFO Point,None,Point,None,Point,None,Point,None,Point
ETF_logo8   _CHAR_INFO Point,None,Point,None,Point,None,Point,None,Point
ETF_logo9   _CHAR_INFO Point,None,None,None,Point,None,None,None,Point
ETF_logo10  _CHAR_INFO Point,None,None,None,Point,None,None,None,Point
ETF_logo11  _CHAR_INFO Point,None,None,None,None,None,None,None,Point
ETF_logo12  _CHAR_INFO Point,Point,Point,Point,Point,Point,Point,Point,Point
```

Fig. 8. School of Electrical engineering, University of Belgrade, bitmap.

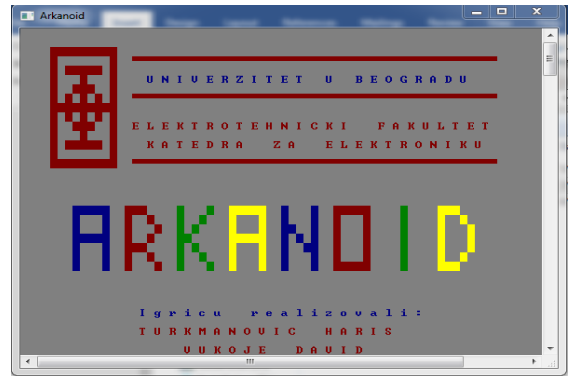

Fig. 9. Welcome window interface

represents coordinates of upper-left and lower-right corner of one block. Information about area position is very important because it's used while detecting block's collision with the ball and while drawing a block on the console. Some blocks have the ability to fall when they are hit twice by the ball. To save information how many times ball hit blocks, CELL structure has a field named "HIT COUNT"

*Pad* is similar to the block and it is also CELL object. The difference is that the pad moves according to the player's wish.

The *Ball* is a different than block's and pad's structure. The ball is the object of the structure named _BALL which has only three fields. Two of them are important for ball movement. The third field is used to store information about the game activity.

## V. GAME ALGORITHMS

In order to introduce better the game algorithms, it is necessary first to explain the way in which the main procedure works. In the Fig. 10 is illustrated program flow graph.

The main procedure should in real-time process information from users and, based on the current graphic object data, determine the next state of the graphic objects. Therefore, it is necessary to have a loop in the main procedure. After the main procedure invokes procedure indebted for creating "Welcome user window", the main procedure goes into a loop which updates graphic object's position. It comes out from that loop if the player loses all his lives or if during the game player is pressed ESC keyboard button. Main procedure loop needs to *write game screen* if the user loses a life, *update pad and ball position* and *update info part* (number of lives and number of points).

Behind the *write game screen* process are located simple algorithms which create a matrix of blocks, pad and a ball. Pad is always located at the center of the bottom console line. The ball is located at the top of the pad and always starts moving away from the pad at the same angle, but the ball's position on the pad is not always same. For that purpose, we used random number generator from the *irvine32* library which produces output in the range from 0 to 1. That output is mull by pad size and adds on pad start position. On that way, we got random ball position on the pad. This algorithm makes the game more interesting and unpredictable.

*Update pad and ball position* procedures have a task to move ball and pad. The algorithm for moving the ball provides the next position of the ball based on the current and previous ball position. Current and previous ball position information are fields of the _BALL structure.

Each time when the ball is moved on the console screen, collision detection algorithms check if a collision has occurred between ball and block, ball and screen corner or ball and the bottom of the screen. If the collision between block and ball is detected, a field named "Hit count" in CELL structure is decremented by 1. If "Hit count" value reaches zero, it informs block update procedures to delete that block from the screen. When a block is deleted from the screen, "Active" field of block structure is set to 0.

After the ball hits any other object or a corner of the screen, ball rebounds at the same as entering angle. Rebound algorithm is based on the simple idea and it is implemented in a procedure named *ReboundBall*. This procedure has arguments which give rebound direction. The previous and current position of the ball are given in *XY* coordinate system, so it's easy to calculate the next ball position according to the given direction based on current and previous directions. Current and previous directions can be used to get information about the previous ball moving direction. This is a very important procedure which is combined with the *FindCollision* procedure.

When next ball position is calculated, *FindCollision* procedure is invoked, because we need to detect if there is a collision, between the ball and some other object, at the next ball position. If we look more closely at the graphic objects, from which the ball can be rebounded, we will see that they have four edges. Because of that, it is important to detect which side of the object the ball hit. Find collision procedure has a task to detect a collision. If a collision is not detected, next ball position is simple to calculate (just keep the current direction of movement). If a collision is detected, this procedure returns an information about which side of the object the ball hit. They are four possible return solutions: *a top side*, *a bottom side*, *left side*, *right side*. With this information and with the previous moving direction of the ball it is possible to rebound ball in the proper direction.

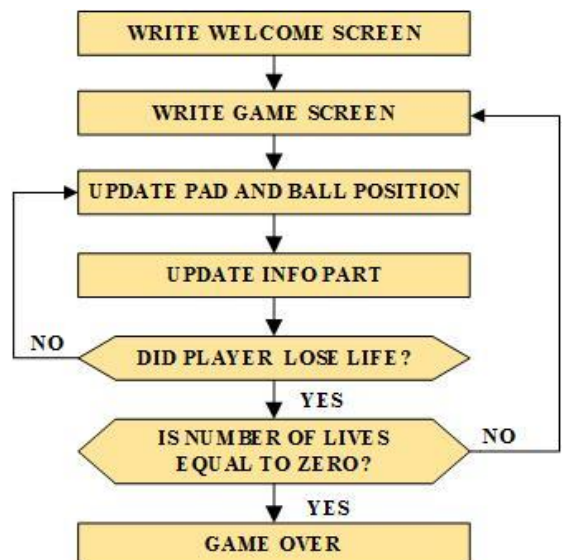

Fig. 10. Main procedure flow char.

Previously listed algorithms for ball rebounding and for finding collision are the most important part of the code and the most complex algorithms in the game. However, there is one more interesting algorithm which allows moving the object to have different speeds of movement.

All "Moving procedure" which implements algorithms for moving objects has one important argument named "Time vector". That is one 16bit word, which keeps the information on how many times the procedure should be called to perform. Thus, the different speed of objects moving is achieved by skipping the procedures calling in the main loop a certain number of times assigned to the "Time vector". In this way, it is possible to get the blocks moving slower than the pad, which makes the game more meaningful.

## VI. CONCLUSION

Purpose of a project named "Arkanoid" was to show the interesting sides of x86 assembly language and to introduce some new and interesting algorithms and ideas which can be implemented in the world of embedded systems. All concepts used by this project (like bitmap, detect object collision, etc.) can be easily implemented on some microcontroller or FPGA. Also, through this project are explained some basic ideas which are hidden behind accessing hardware through interface functions. So, this paper can be interesting either from software or from a hardware point of view. This project is interesting for the educational purposes and it can be used as a interesting approach for teaching assembly programming.

During the work on this project, main gold was to make a more flexible project that can be upgraded in the future. From this point of view, next upgrade of this project will be a user-friendly game menu which will allow the gamer to select desired mode of the game.

## VII. ACKNOWLEDGMENT

The work has been supported through the Republic of Serbia Ministry of Science project TR33020.

## VIII. REFERENCES

[1] K. R. Irvine and L. B. Das, *Assembly language for x86 processors*: Prentice Hall, 2011.
[2] M. Prokin, *Računarska elektronika*. Belgrade: Akademska misao, 2006.
[3] K. R. Irvine, *Assembly language for Intel-based computers*: Citeseer, 2003.
[4] D. D. Gajski, S. Abdi, A. Gerstlauer, and G. Schirner, *Embedded system design: modeling, synthesis and verification*: Springer Science & Business Media, 2009.
[5] M. Wolf, *High-performance embedded computing: applications in cyber-physical systems and mobile computing*: Newnes, 2014.
[6] R. Zurawski, *Embedded systems handbook*: CRC press, 2005.
[7] S. M. Kuo, B. H. Lee, and W. Tian, *Real-time digital signal processing: fundamentals, implementations and applications*: John Wiley & Sons, 2013.
[8] B. Stroustrup, *The C++ programming language*: Pearson Education India, 2000.